\newif\ifanonymousversion
\anonymousversionfalse

\ifanonymousversion

\documentclass[compsoc,conference,a4paper,10pt,times]{IEEEtran}

\else

\documentclass[conference]{IEEEtran}

\fi

\usepackage{tikz}
\usepackage{amsmath}
\usepackage{xcolor}
\usepackage{hyperref}
\usepackage{soul}
\usepackage[braket, qm]{qcircuit}
\usepackage{physics}
\usepackage{enumitem}
\usepackage{multirow}
\usepackage{booktabs}
\usepackage{algorithmicx}
\usepackage{algpseudocode}
\usepackage{algorithm}
\usepackage{balance}
\usepackage{nicematrix}
\usepackage{subcaption}
\usepackage{amsfonts}
\usepackage{flushend}

\newcommand{\nsf}[1]{\href{https://www.nsf.gov/awardsearch/showAward?AWD_ID=#1}{#1}}

\begin{document}

\date{}

\title{Research Directions\\ in Quantum Computer Cybersecurity}

\ifanonymousversion

\else

\author{
\IEEEauthorblockN{Jakub Szefer}
\IEEEauthorblockA{\textit{
Dept. of Electrical \& Computer Engineering} \\
\textit{Northwestern University}\\
Evanston, IL, USA \\
jakub.szefer@northwestern.edu}
}

\fi

\ifanonymousversion

\else

\IEEEoverridecommandlockouts
\makeatletter\def\@IEEEpubidpullup{6.5\baselineskip}\makeatother
\IEEEpubid{\parbox{\columnwidth}{
This work was supported in part by NSF grant \nsf{2332406}.
}
\hspace{\columnsep}\makebox[\columnwidth]{}}

\fi

\maketitle
\pagestyle{plain}

\begin{abstract}
This document presents a concise overview of the contemporary research directions in quantum computer cybersecurity. The aim of this document is not to be a survey, but rather a succinct summary of the major research directions in quantum computer cybersecurity at the end of the first half of the current decade. The document has been inspired by the presentations and discussions held at the 3$^{rd}$ Quantum Computer Cybersecurity Symposium, but goes beyond the contents of the symposium and aims to summarize at the high level the last five years of quantum computer cybersecurity work in academia. It is hoped that the document can provide researchers as well as government and industry leaders an overview of the current landscape of security threats and defenses against emergent quantum computing technologies. The document also includes a discussion of the current trends in cybersecurity research on quantum computers, and the perceived research gaps that should be filled with future funding and through academic and industry~research.
\end{abstract}

\section{Introduction}

Quantum computing research and deployment of (small, but growing in size) quantum computers have expanded rapidly in the last decades. We have seen quantum computers go from $2$ qubits~\cite{hidary2021brief} in late $1990$s, to one of the first cloud-based IBM quantum computers with a few qubits in around $2016$, to now machines with $100$s of qubits and projections of having orders of magnitude larger machines by the end of this decade~\cite{ibmqubit}. The quantum computing ecosystem is now huge with not only academia contributing, but with countless startups and established industry companies working on all aspects of quantum computing: from software and algorithms, through compilers, to hardware. As of the end of $2025$, there is a large and growing number of cloud-based services where users can access real quantum computers. Some major cloud providers that offer quantum computers include IBM Quantum~\cite{ibm_quantum}, Amazon Braket~\cite{braket}, and Microsoft Azure~\cite{azure}. But the list is growing each year, and many quantum computer vendors now also offer their own cloud services.

Research on cybersecurity of quantum computers is much younger than that on quantum computing in general. AA rough but useful point of reference is the start of the current decade when significant interest in quantum computer cybersecurity has emerged. Since about $2020$, numerous research papers have emerged showing various theoretical and proof-of-concept demonstrations of potential attacks on quantum computers. This growth in quantum computer cybersecurity was driven from two directions: by security researchers who started to work on quantum computing, and in a smaller part by quantum computer researchers who saw an opportunity for new ways of thinking about quantum computers by considering the security of these machines.

Discourse about quantum computer cybersecurity is also reaching broader audiences and more and more academic venues have emerged where researchers can share their findings. The 1$^{st}$ Quantum Computer Cybersecurity Symposium (QCCS) was held in $2023$, for example, and many quantum computer security-related workshops and special sessions at hardware, architecture, and quantum computing conferences have emerged since then. Through these venues, has begun to be more visible and to reach more researchers in academia and industry. Thanks to events such as the Quantum Computer Cybersecurity Symposium, more attention is now being paid to the security issues (and possible defenses) of quantum computers. Most recently, the 3$^{rd}$ Quantum Computer Cybersecurity Symposium attracted the largest interest from academia and industry, with the highest level of sponsorship and participation. 

Quantum computer cybersecurity has some natural overlaps with post-quantum cryptography (PQC), but is actually an orthogonal research direction. It is useful to clarify this distinction as the mention of ``security'' together with ``quantum computing'' is often interpreted as meaning something related to post-quantum cryptography. This document, and quantum computer cybersecurity research discussed here, do not cover topics related to PQC. While post-quantum cryptography deals with protection of classical computers, networks, and data from attackers who have access to quantum computers (and thus may break some of the current cryptographic algorithms), quantum computer cybersecurity deals with protection of quantum computers themselves from security attacks. Quantum computer cybersecurity also deals with protection of users' quantum circuits or programs, including intellectual property of the algorithms and the circuits that realize them, and any data processed by quantum computers. It further extends to protection of quantum computers from attacks, especially considering the now omnipresent cloud-based quantum computers that almost anybody can access (and possibly~attack).

\section{Document Organization}

This document provides an overview of the current research directions in quantum computer cybersecurity. The document is not a survey and has limited citations. It instead focuses on presenting a high-level summary of the key research directions, and research gaps that researchers and government and industry leaders should be aware of when they consider quantum computer security attacks and defenses.

A detailed, but naturally incomplete, listing of quantum computer cybersecurity research papers can be found in the public GitHub repository\footnote{BibTeX Repository for Quantum Computer
Cybersecurity Research, \url{https://github.com/caslab-code/qc-hardware-cybersecurity-bibtex}} maintained by the author~\cite{githubbibtex}. The repository contains BibTeX files that are regularly updated with new research papers that appear at the various conferences and on arXiv. The repository can be considered a companion to this document. Perspectives and research directions presented in this document are drawn from the research papers contained in the repository.

\section{Security in the NISQ Era}

Quantum computing can be broadly divided into NISQ (Noisy Intermediate-Scale Quantum) and FTQC (Fault Tolerant Quantum Computing) regimes. Much of the early and current cybersecurity research on quantum computers focuses on NISQ, which is covered in this section.

NISQ quantum computers are generally described as machines with a small number of qubits, with relatively higher error rates, and no error correction. NISQ quantum computers are thus very noisy and unable to provide real benefits in many domains, although there is hope and promise of having benefits when using NISQ quantum computers with variational algorithms, for example. From a security perspective, perhaps the most interesting aspect of NISQ quantum computers is that many of them are available as cloud-based services that anybody can~access, opening up possibilities of interesting security threats. In parallel, quantum computers are moving out of research labs, and can be found in hospitals or university campuses where they are outside strictly secure data center settings. This opens up other security threats, especially due to physical attacks.

\subsection{Major NISQ Quantum Computer Cybersecurity Research}

Among NISQ quantum computer cybersecurity research, a strong theme is that researchers have focused on exploiting the noisy nature of NISQ machines to demonstrate various proof-of-concept attacks. It should be noted that in many cases these are not yet real attacks as they make certain assumptions, such as multi-tenancy, which is not currently offered by quantum computer providers. Nevertheless, the research demonstrates in many cases that quantum computers are connected to the cloud with almost no security mechanisms in place and are quite~vulnerable to attacks.

The majority of existing research on NISQ quantum computer cybersecurity falls into the following categories:

\begin{enumerate}
    \item \textbf{Attacks Exploiting Gate Crosstalk}: Perhaps one of the first and major targets of security researchers working on security of quantum computing was gate crosstalk. Due to the noisy nature of NISQ quantum computers, and lack of error correction, as well as use of fixed frequency couplings (in superconducting qubit quantum computers), it has been shown that it is relatively easy for an ``attacker'' circuit with access to some qubits to induce noise onto a ``victim'' circuit running concurrently on the same machine. Such gate crosstalk can be achieved by the attacker executing quantum gates on its assigned qubits as the victim is doing some computation on its assigned qubits. This can cause the victim circuit to generate wrong or incorrect results: a type of attack on the integrity of the computation. These attacks assume multi-tenancy, which is not yet offered by quantum computer providers. This is however by far the largest area of security focus in the last years as noisy gates and qubits are vulnerable to disruptions and researchers were able to show various changes in victim circuit outcomes. Limited work has also explored the crosstalk in the reverse direction where attacker attempts to observe what computation the victim is doing by measuring the crosstalk that victim's computation induces on attacker's qubits. This has been explored only in limited cases, but demonstrates a type of attack on confidentiality of the victim's computation. Most work has been done on superconducting quantum computers, especially IBM, in large part thanks to IBM's willingness to make the machines available, in many cases for free, to researchers. Recent advances in tunable couplers and emergent fault-tolerant quantum computers may render these types of attacks obsolete, but nevertheless, research on these crosstalk attacks continues and shows new and novel results every few months.
    
    \item \textbf{Attacks Exploiting Reset Gates}: An interesting class of attacks has focused on reset gates in superconducting qubit quantum computers. Imperfect operation of the reset gates was abused to show that an attacker circuit could induce incorrect initial qubit states (by manipulating qubits with higher energy states which are not reset correctly by reset gates) so that the victim circuit that runs subsequently generates incorrect computation results. And, it was shown that attacker circuits running after victim circuits could potentially steal the outputs of the computation. The reset gate attacks could violate both integrity and confidentiality of the victim circuit's computation. They also do not require multi-tenancy. However, they do assume use of active reset gates between shots of the victim and attacker circuits, and that victim and attacker will be interleaved at the granularity of shots. It is not clear if this setting is currently supported by quantum computer providers as scheduling algorithms and use of active reset between different users' circuits is not documented publicly.
    
    \item \textbf{Attacks Exploiting Readout Crosstalk}: One of the newest categories of security attacks in quantum computers leverages readout crosstalk. A feature of superconducting quantum computers is that many qubits may share a single readout line. During readout, the output or state of one qubit may affect the readout of another, thus leading to a confidentiality attack (if the attacker sets a qubit into some state and then tries to observe how the victim's readout affected it) or an integrity attack (if the attacker sets one of their qubits into a specific state that is likely to affect the readout observed by the victim). This type of attack has a fairly constrained threat model, however, as it requires both multi-tenancy and the victim and the attacker have to access qubits that share the same readout lines. The attacks, nevertheless, point out an important lesson that microarchitecture of the quantum computers (e.g., how qubits share readout lines) can have significant security~implications.
    
\end{enumerate}

\noindent The various research described above, which again can be found in papers referenced in the quantum computer cybersecurity BibTeX repository~\cite{githubbibtex}, has disproportionately focused on superconducting quantum computers, and mainly from IBM. This is due to early and easy access to IBM quantum computers. However, works focusing on other types of quantum computers and other providers do exist and may become more common as other providers give more access to the quantum computers through cloud services, and as free IBM access is reduced and queue times increased on IBM. Note that security of the software stack and the hardware controllers is discussed separately later in this document. 

Many of the attacks have so far been demonstrated on real quantum computers, thanks to free quantum credits or access to the machines, and ability to use various features such as pulse-level access. Casually it can be observed, however, that these features and free access are getting harder to obtain. As quantum computing continues to explode in popularity, free access is reduced, and non-security research projects seem to get priority when requesting quantum credits or free access. Further, access to data such as qubit calibration data, qubit frequencies, or details of the quantum chips is being reduced, making it harder to analyze the machines. Other features such as pulse-level access have also been eliminated by a number of quantum computer providers in recent months. As the features and access is reduced, the future years may see more of reverse-engineering type security research, not different in spirit from security research on classical processors where various attacks on processor caches or speculative execution have been found by experimenting with and reverse-engineering processor features~\cite{szefer2018principles}.

\section{Security in the FTQC Era}

As work on NISQ quantum computers progresses, quantum computing is steadily advancing into the fault-tolerant regime. The second half of this decade will almost certainly be about FTQC quantum computers. FTQC quantum computers are characterized by use of error correction, where many physical qubits are used to realize one virtual qubit. This requires many physical qubits, but has the advantage of almost error-free virtual qubits that can be used to achieve quantum advantage on many algorithms, such as Shor's, that really only work on error-free quantum computers. FTQC introduces new features into the quantum computer, notably the error correction codes and the software and hardware used to decode and correct the errors. FTQC also abstracts away and reduces the access that users have to the qubits as each user is given access to a set of logical qubits and in ideal setting knows nothing about the physical realization of the hardware. This is not different from the instruction set architectures of classical processors where users know instructions, registers, and memory that the computer supports, but the microarchitectural realization of the processor is hidden away. Such abstraction in classical computers did not prevent security threats~\cite{szefer2019survey}, and same is likely for FTQC.

As FTQC quantum computers are only now emerging, security research on FTQC quantum computers is also just emerging. It should be noted that FTQC quantum computer access is not available to general public via cybersecurity of the cloud services. Thus unlike NISQ security research that in many cases could demonstrate some experiments on real quantum computers, FTQC security research so far is limited to simulations of the attacks on simulated systems.

\subsection{Major FTQC Quantum Computer Cybersecurity Research}

A new and different feature of FTQC quantum computers is the use of the error correction codes. Consequently it is natural that the security research in the FTQC domain focuses around the error correction.

The majority of emerging research on FTQC quantum computer
cybersecurity falls into the following categories:

\begin{enumerate}
    \item \textbf{Attacks on Error Correction}: Limited but emerging work is focusing on error correction in the adversarial regime. Error correction has not been analyzed well from the perspectives of attackers who may want to manipulate the error correction to cause the logical qubits to fail to error correct, thus forming a type of integrity attack on the computation. Finding ways that physical qubits could be flipped is interesting and future direction that many will likely explore. On the other hand, with limited access only to logical qubits, it remains an important open question how attackers could trigger errors in the physical qubits if they only can control logical qubits. Physical attacks may be one way around this limitation.
    
    \item \textbf{Attack on Decoders}: FTQC quantum computers require expanded decoders, typically part of the quantum computer controller, that deal with the error correction operations. Emerging research has already started to explore security from the perspective of the decoders and how the error corrected computation progresses. Emergent work has demonstrated the ability to observe the (logical) quantum gate operations from the traces in FTQC controllers: a type of confidentiality attack. Other work has explored ways to modify operation of the controllers or the decoders to induce faults in the qubits: a type of integrity attack. As quantum controllers are central to FTQC, more security attacks will likely emerge in this area.
    
\end{enumerate}

\noindent The recent increase in security work on FTQC demonstrates that security research is closely following evolution of FTQC. It is difficult to observe trends in FTQC cybersecurity work yet, but the focus on the error correction codes and decoders is clear.

An interesting inflection point will be once FTQC machines are available for users to actually use. Currently many FTQC advances are presented in academia and news releases, but these machines are not easily accessed nor available. There may be a more quiet period as researchers await access to real FTQC machines for testing. More simulation-based and theoretical attacks may emerge in the meantime.

\section{Security Across NISQ and FTQC Eras}

Both NISQ and FTQC quantum computers share many features, and cybersecurity research of these can affect both NISQ and FTQC machines. Attacks and research that touches both NISQ and FTQC include:

\begin{enumerate}
    \item \textbf{Attacks on Software and Software Supply Chain}: Software plays key role in quantum computing. From various software development kits, through compilers, to public repositories of sample algorithms and code libraries, a huge amount of quantum computing is software focused. Software supply chain security has been largely not yet explored in quantum computing setting. Some CVEs have been reported for software development kits, but much work will likely emerge on security attacks on the~software.
    
    \item \textbf{Attack on Hardware and Hardware Supply Chain}: Hardware is tightly controlled in quantum computing world. Vendors are said to physically shred unused and old quantum computing chips to prevent leakage of any intellectual property. In parallel, the actual set of hardware suppliers is very limited, making it easy to control the hardware. As quantum computing expands and becomes commoditized, keeping track of the hardware suppliers and vendors will be harder. New types of supply chain attacks may emerge, and the trust in the existing hardware may be reduced. Early work on quantum computer fingerprinting may gain more attention again; although intersection of fingerprinting and limited access to hardware due to FTQC may create interesting challenges for identifying and tracking hardware.
    
\end{enumerate}

\noindent The work on security of software and hardware has many overlaps with classical security research. Adapting known techniques may be key to quickly finding new attacks, and then provisioning security features for defense.

\section{Emerging Research on Defenses for Quantum Computers}

A theme is clear from the discussion in the prior sections: much of the current work on quantum computer cybersecurity focuses on attacks. Attacks usually gain more attention and tend, at least initially, be easy to publish in academic venues. Work on defenses has higher bar to pass. Defenses almost always present some overhead. While this is natural, almost any degradation in quantum computer fidelity, or increase in circuit size or duration due to some defense is difficult to accept. In some sense this is not unlike in classical computers. Performance (here fidelity) is easy to quantify, while security is not. Thus any gains in security are difficult to quantify, making it hard to measure the security benefit vs. performance~overhead that they induce.

Nevertheless a body of work has emerged on defenses. From early work on detecting malicious quantum circuits to security architectures for quantum computers, researchers are working on designing defenses. A unique opportunity, which was missed with classical computers, is that the defenses could be integrated now, before quantum computers are commoditized, ensuring that security is built into quantum computers from the beginning. As more work on defenses is presented the research balance may shift from attacks to defenses.

\section{Conclusion}

The second half of the current decade should bring about expansion in the security of quantum computers, especially focusing on FTQC. Many research gaps remain that will hopefully be filled in by research funding that supports new research discoveries. In no particular order, and necessarily incomplete, a list of research gaps includes: 1) verification and protection of the quantum computing software and SDKs from attacks, 2) passive and active means to analyze user circuits for any malicious activity, 3) protection of user circuits from malicious compilers and SDKs, 4) security architectures to ensure integrity and confidentiality of user's code as it runs on remote quantum computers, 5) physical security as quantum computers move out of secure data centers, 6) validation and fingerprinting of quantum computing hardware and proof that code was executed on a quantum computer, 7) anti reverse engineering and protections for quantum hardware vendors and their intellectual property, and 8) security of quantum networking, quantum sensing, quantum memories and any other technologies that get integrated and connected to quantum computers. The future research will hopefully address these and other emerging threats to quantum computation. The future will hopefully also bring about increase in the importance of security features for quantum computation and that software and hardware vendors prioritize not only performance and fidelity, but also security of their quantum computers.

\section*{Acknowledgment}

This document has been especially inspired by the presentations and discussions held at the 3$^{rd}$ Quantum Computer Cybersecurity Symposium\footnote{Quantum Computer
Cybersecurity Symposium (QCCS '25), \url{https://caslab.io/events/qccs}}. It is also based on the various research findings published by the academic community and captured in the public BibTeX repository~\cite{githubbibtex}.

\balance

\bibliographystyle{plain}
\bibliography{bibliography.bib}

\end{document}